\documentclass{IEEEtran4PSCC}
\ifCLASSINFOpdf
\usepackage[pdftex]{graphicx}
\graphicspath{{figures/}}
\else
\usepackage[dvips]{graphicx}
\graphicspath{{figures/}}
\fi
\usepackage{tikz}
\usetikzlibrary{shapes.geometric, arrows,positioning}

\usepackage[cmex10]{amsmath}
\usepackage{amssymb}
\ifCLASSOPTIONcompsoc
  \usepackage[caption=false,font=normalsize,labelfont=sf,textfont=sf]{subfig}
\else
  \usepackage[caption=false,font=footnotesize]{subfig}
\fi
\usepackage{multirow}
\usepackage{booktabs}

 \usepackage{url}

\usepackage{cleveref}

\usepackage{xspace}
\usepackage[super]{nth}

\newcommand{\tf}{\mathrm} 

\crefname{equation}{}{}
\Crefname{equation}{Equation}{Equations}
\creflabelformat{equation}{#2(#1)#3}

\crefrangelabelformat{equation}{(#3#1#4)-(#5#2#6)}

\crefname{figure}{Fig.}{Fig.}
\Crefname{figure}{Fig.}{Fig.}
\creflabelformat{figure}{#2#1#3}

\crefname{table}{Table}{Table}
\Crefname{table}{Table}{Table}
\creflabelformat{table}{#2#1#3}

\crefname{section}{Section}{Sections}
\Crefname{section}{Section}{Sections}
\creflabelformat{section}{#2#1#3}

\newcommand{\ts}{\ensuremath{TS}}
\newcommand{\pv}{\text{PV}\xspace}

\newcommand{\unit}[1]{\,$\mathrm{#1}$\xspace}
\newcommand{\ctskwh}{\unit{\frac{cts}{kWh}}}
\newcommand{\kchfmva}{\unit{\frac{kCHF}{MVA}}}
\newcommand{\pload}{\ensuremath{P^\tf{load}}}
\newcommand{\ppv}{\ensuremath{P^\pv}}

\newcommand{\pcur}{\ensuremath{P^\tf{cur}}}
\newcommand{\pbat}{\ensuremath{P^\tf{bat}}}

\newcommand{\pvcap}{\ensuremath{P^{\pv}_\tf{cap}}}
\newcommand{\ebatcap}{\ensuremath{E^\tf{bat}_\tf{cap}}}

\newcommand{\sit}{\ensuremath{{i,t}}}
\newcommand{\skt}{\ensuremath{{k,t}}}
\newcommand{\sikt}{\ensuremath{{ik,t}}}

\newcommand{\cimp}{\ensuremath{c^\tf{imp}}\xspace} 
\newcommand{\cexp}{\ensuremath{c^\tf{exp}}\xspace} 

\newcommand{\opex}{\emph{OPEX}\xspace}

\newcommand{\cbat}{\ensuremath{c^\tf{bat}}} 
\newcommand{\cfbat}{\ensuremath{c^\tf{bat}_F}} 

\makeatletter
\let\old@ps@headings\ps@headings
\let\old@ps@IEEEtitlepagestyle\ps@IEEEtitlepagestyle
\def\psccfooter#1{%
	\def\ps@headings{%
		\old@ps@headings%
		\def\@oddfoot{\strut\hfill#1\hfill\strut}%
		\def\@evenfoot{\strut\hfill#1\hfill\strut}%
	}%
	\def\ps@IEEEtitlepagestyle{%
		\old@ps@IEEEtitlepagestyle%
		\def\@oddfoot{\strut\hfill#1\hfill\strut}%
		\def\@evenfoot{\strut\hfill#1\hfill\strut}%
	}%
	\ps@headings%
}
\makeatother

\psccfooter{%
	\parbox{\textwidth}{\hrulefill \\ \small{22nd Power Systems Computation Conference} \hfill \begin{minipage}{0.2\textwidth}\centering \vspace*{4pt} \includegraphics[scale=0.06]{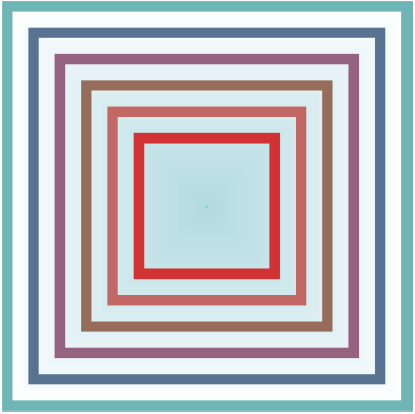}\\\small{PSCC 2022} \end{minipage} \hfill \small{Porto, Portugal --- June 27 -- July 1, 2022}}%
}

\newcommand{\fw}{.8\columnwidth}

\begin{document}
	%
	\title{Distributed flexibility as a cost-effective alternative to grid reinforcement }

	\author{
		\IEEEauthorblockN{Jordan Holweger, Christophe Ballif, Nicolas Wyrsch}
		\IEEEauthorblockA{Photovoltaics and thin film electronics laboratory (PV-LAB) \\
			\'Ecole Polytechnique F\'ed\'erale de Lausanne (EPFL), Institute of Electrical and Micro Engineering (IEM)\\
			Neuch\^atel, Switzerland\\
			\{jordan.holweger, christophe.ballif, nicolas.wyrsch\}@epfl.ch}
	}


	\maketitle
	
	\begin{abstract}
		The deployment of distributed photovoltaics (PV) in low-voltage networks may cause technical issues  such as voltage rises, line ampacity violations, and transformer overloading for distribution system operators (DSOs).  These problems may induce high grid reinforcement costs. In this work, we assume the DSO can control each prosumer's battery and PV system. Under such assumptions, we evaluate the cost of providing flexibility and compare it with grid reinforcement costs. Our results highlight that using distributed flexibility is more profitable than reinforcing a low-voltage network until the PV generation covers 145\% of the network annual energy demand. 
	\end{abstract}
	
	\begin{IEEEkeywords}
		PV, flexibility, grid reinforcement cost, battery, optimal power-flow
	\end{IEEEkeywords}

	
	\section{Introduction}
	The fast deployment of distributed photovoltaics (PV) causes numerous challenges to distribution system operators (DSOs). The imbalance between local generation and load can create technical problems in low-voltage grids, such as line ampacity violations, overvoltage, and transformer overloading \cite{Viral2012}. A DSO must take countermeasures such as grid reinforcement (GR) \cite{Scheidler2017,Vu2018}. Alternatively, future distributed PV systems might provide a significant degree of flexibility to reduce the need for GR. Indeed, battery energy storage systems (BESSs)  \cite{Hashemipour2018}, inverter reactive power capability \cite{Olivier2016,Olivier2018,Prionistis2021}, and active power curtailment (APC) are possible solutions to increase a network's flexibility and mitigate GR costs \cite{Spiliotis2016}. Recent literature demonstrated that BESSs might increase a network's PV hosting capacity but are not competitive with GR \cite{Gupta2021,Gupta2021a}. This work compares the GR cost with the cost of harvesting distributed flexibility, particularly BESS, APC, and inverter reactive power capabilities for various PV penetration levels.

	\section{Notation and basic formulas}
\begin{figure}[!ht]
    \centering
    \includegraphics[width=.5\columnwidth]{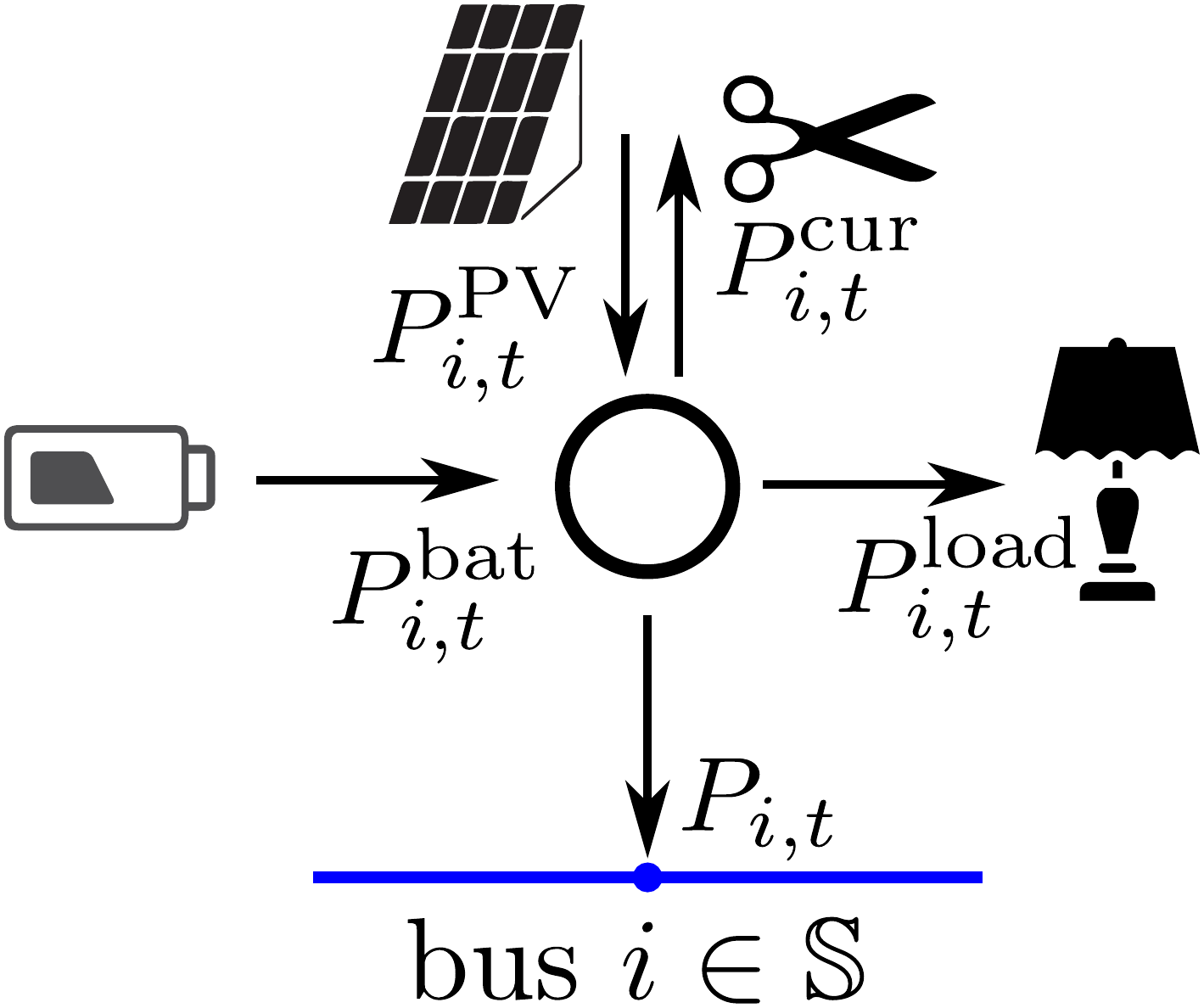}
    \caption{System power balance.}
    \label{fig:pow_bal}
    \end{figure}

	This section aims to present the basic notation and symbols used in this paper. This work analyses a low-voltage grid with connected prosumers with a BESS, which are considered as static loads or generators injecting active and reactive power $P_\sit,Q_\sit$ at their grid connection points. The active power consists of the PV generation $\ppv_\sit$, the active power curtailment $\pcur_\sit$, the BESS's exchange power $\pbat_\sit$ (positive when discharging), and the uncontrollable load $\pload_\sit$. \Cref{fig:pow_bal} illustrates the power balance of each prosumer. 
	
    \subsubsection*{Sets}
	\begin{itemize}
	    \item $\mathbb{B}, \mathbb{L},\mathbb{S}$ sets of buses, lines, and prosumers' systems ($\mathbb{S} \subset \mathbb{B}$) respectively.
	    \item $\mathbb{L}^\tf{tr}$ set of transformer branches (pair of bus $(ik)$ with $i$ the high-voltage bus). 
	\end{itemize}
	
	 \subsubsection*{Parameters}
	\begin{itemize}
	    \item $j=\sqrt{-1}$ imaginary number. 
	    \item $\pload_i$ uncontrollable active power load at bus $i$.
	    \item $\ppv_i$  \pv generation at bus $i$. 
	    \item $G_{ik},B_{ik}$ conductance/susceptance on line $(ik)$.
	    \item $\overline{x},\underline{x}$ upper and lower limit for variable $x$.
	    \item $x^+,x^-$ positive and negative part of $x$.
	    \item $q_r^\tf{max}$ inverter reactive power capabilities ratio (MVAr/MW).
	    \item \cimp, \cexp import and export tariff (consumer perspective).
	    \item $L,r$ component lifetime and interest rate to calculate the annualisation factor $R$.
	\end{itemize}
	
	 \subsubsection*{Variables}
	\begin{itemize}
	    \item $\pbat_{i,t}$, battery active power (positive when discharging).
	    \item $\pcur_{i,t}$, PV active power curtailment  uncontrollable active power load at bus $i$.
	    \item $Q^{\pv}_{i,t}$ PV inverter reactive power injection.
	    \item $S_i^\tf{tr}$ transformer apparent power. 
	    \item $V_{i,t}$, $I_{ik,t}$ voltage magnitude at bus $i$ and current in line $(ik)$ and time $t$.
	    \item $\hat{\theta}_\sikt$ voltage angle difference between bus $i$ and $k$ at time $t$.
	\end{itemize}
	
	 \subsubsection*{AC load flow equations}
	
	\begin{subequations}
	    \begin{align}
	        \label{eq:P1}
	        P_\sit &= \begin{cases}
	        \ppv_\sit - \pcur_\sit +\pbat_\sit -\pload_\sit & i\in \mathbb{S}\\
	        0 & i \in \mathbb{B}- \mathbb{S}
	        \end{cases}\\
	        \label{eq:Q1}
	        Q_\sit &= \begin{cases}
	        Q^{\pv}_\sit & i\in \mathbb{S}\\
	        0 & i \in \mathbb{B}- \mathbb{S}
	        \end{cases}\\
	        \label{eq:P2}
	        P_\sit &= V_\sit\sum_{k\in \mathbb{B}} V_\skt\left(G_{ik}\cos{\hat{\theta}_\sikt } + B_{ik}\sin{\hat{\theta}_\sikt } \right)\\
	        \label{eq:Q2}
	        Q_\sit &= V_\sit\sum_{k\in \mathbb{B}} V_\skt\left(G_{ik}\sin{\hat{\theta}_\sikt } - B_{ik}\cos{\hat{\theta}_\sikt } \right)\\
	        \label{eq:I}
	        I_\sikt &= \left(G_{ik} + jB_{ik}\right)\cdot \left(V_\sit - V_\skt\right)
	    \end{align}
	\end{subequations}
	
	\subsubsection*{Network operation constraints}
	\begin{subequations}
	    \begin{align}
	        \label{eq:Vbounds}
	        \underline{V} \leq V_\sit \leq&  \overline{V}   &&\forall \quad i\in \mathbb{B}\\
	        \label{eq:Ibounds}
	        I_\sikt \leq& \overline{I}  & &\forall \quad (ik) \in \mathbb{L}\\
	        \label{eq:TRbounds}
	        S^\tf{tr}_{i}=\left| V_\sit \cdot I_\sikt \right| \leq& \overline{S}^\tf{tr}_{i}  & &\forall \quad (ik) \in \mathbb{L}^\tf{tr}
	    \end{align}
	\end{subequations}
	
	\section{Methodology}
    The methodology is based on a sequential approach. First, following the method described in \cite{bloch_impact_2019}, the BESS's capacity \ebatcap and its optimal control trajectory are optimised for each prosumer  with a fixed PV penetration (fixed PV capacity for each system, \pvcap). The optimisation problem aims to minimise the total cost of ownership of each BESS-PV system  \cref{eq:min_totex} composed of the sum of the annualised BESS capital cost and the system operating cost \cref{eq:opex0},  subject to the power balance \cref{eq:P1}. The operating costs \cref{eq:opex0} are the cost of exchanging energy with the grid under an import and export tariff (from a prosumer's perspective) \cimp and \cexp, respectively. As the PV capacity is fixed, it does not appear in the objective function. The term $\sigma$ \cref{eq:sigma} is a regularisation cost whose aim is to minimise battery usage. The annualisation factor $R$ depends on the considered lifetime $L$ and interest rate $r$. More details about the PV and BESS model can be found in \cite{bloch_impact_2019}. 
    
    Second, the load flow problem \crefrange{eq:P1}{eq:I} is solved  at each point in time to calculate the network variable states, $V_\sit$, $I_\sikt$, and $ S^\tf{tr}_{i}$. Then, the annualised cost of GR \cref{eq:reinf} is calculated as the sum of the cost of replacing lines \cref{eq:reinf_line} because of line ampacity violation \cref{eq:bin_line}, and the cost of replacing transformers due to an apparent power limit violation \cref{eq:bin_trafo}. The annualisation factor $R^\tf{grid}$ is calculated as in \cref{eq:bat_ann}. The cost of replacing a line is proportional to the line length $d_l$ and its specific cost per unit of length $c^\tf{line}$. The cost of replacing a transformer is simply the product of the transformer cost per unit of capacity $c^\tf{trafo}$ and the new maximum apparent power $\max_t S^\tf{tr}_t$.
    
    Third, distinct time domains $t\in \Pi_m,\, m=1\ldots M$ are constructed in which network constraints \crefrange{eq:Vbounds}{eq:TRbounds} are not satisfied. These contiguous time steps represent intervention periods in which the DSO must undertake actions to prevent violations of network constraints. To model these actions, we solve an optimal power-flow problem for each period $\Pi_m$  with the goal of minimizing the PV curtailed energy \cref{eq:flex_opf_opt}, which determines where and when to charge or discharge the batteries, inject or absorb reactive power (within the inverter capability, $q_r^\tf{max}$ constraints \cref{eq:q_max}), and perform APC. Due to the multi-period nature of this problem, any action altering the battery state of the charge profile may have consequences in the future. To prevent such distortion, the BESS's energy content at the beginning and end of the intervention period should match the original energy content from the optimal control trajectory \cref{eq:soc_constr}. We implemented this constraint in the \textsc{PowerModels} \cite{powermodels} library that we used to solve the optimal power-flow problem. Note that the BESS model is slightly different in \textsc{PowerModels} than in \cite{bloch_impact_2019}. Typically in the former implementation, we neglected the active power losses, did not consider injection impedance, and assumed that the battery cannot provide reactive power. 
    
\begin{figure}[t]
    \centering
    \newcommand{\bwidth}{.4\columnwidth}
\newcommand{\vnodespace}{.05\columnwidth}
\newcommand{\vvnodespace}{.07\columnwidth}
\newcommand{\hnodespace}{.07\columnwidth}
\newcommand{\bheight}{1.2cm}
\tikzstyle{box} = [rectangle, draw=black, thick,minimum width=\bwidth, minimum height=\bheight,text centered,text width=\bwidth]
\tikzstyle{decision} = [diamond, draw,minimum width=3cm, minimum height=1cm, text centered, draw=black, thick,fill=white]
\tikzstyle{arrow} = [thick,->,>=stealth]

\begin{tikzpicture}[node distance=.15\columnwidth,auto,font=\small]

\node (start) [box]  {\textbf{Start}};
\node (end) [box,right=\hnodespace of start] {\textbf{End}};
\node (s1) [box,below=\vvnodespace of start] {PV scenario $s=1$};
\node (s2) [box,below=\vvnodespace of s1] {Solve optimal BESS capacity and control trajectory \crefrange{eq:min_totex}{eq:bat_ann}};
\node (s3) [box,below=\vvnodespace of s2] {Solve load flow equations \crefrange{eq:P1}{eq:I}};
\node (s4) [box,below=\vvnodespace of s3] {Evaluate  GR cost \crefrange{eq:reinf}{eq:reinf_trafo}};
\node (s5) [box,below=\vvnodespace of s4] {Find set of continuous intervention periods $t \in \Pi_m \leftrightarrow $ \crefrange{eq:Vbounds}{eq:TRbounds} not satisfied $m=1 \ldots M$};
\node (s6) [box,below=\vvnodespace of s5] { First intervention period 
$m=1$
Initialize grid exchange $P^\prime_{i,t} = P_{i,t} \quad \forall i,t $};
\node (s7) [box,right=\hnodespace of s6] {Solve optimal power flow problem \cref{eq:flex_opf_opt} s.t. \cref{eq:P1,eq:Q1,eq:P2,eq:Q2,eq:I,eq:Vbounds,eq:Ibounds,eq:TRbounds,eq:soc_constr}, update $P^\prime_{i,t} \quad \forall i, t\in \Pi_m$};
\node (s8) [box,above=\vnodespace of s7] {$ m=m+1$};
\node (d1) [decision,above=\vnodespace of s8] {$m>M$};
\node (s9) [box,above=\vvnodespace of d1] {Evaluate cost of providing flexibility \cref{eq:flex_cost}};
\node (s10) [box,above=\vnodespace of s9] {$s=s+1$};
\node (d2) [decision,above=\vnodespace of s10] {$s>S$};

\draw [arrow] (start) -- (s1);
\draw [arrow] (s1) -- (s2);
\draw [arrow] (s2) -- (s3);
\draw [arrow] (s3) -- (s4);
\draw [arrow] (s4) -- (s5);
\draw [arrow] (s5) -- (s6);
\draw [arrow] (s6) -- (s7);
\draw [arrow] (s7) -- (s8);
\draw [arrow] (s8) -- (d1);
\draw [arrow] (d1) -- node[anchor=north west] {yes} (s9);
\draw [arrow] (d1.east)  node[anchor=south ] {no} -- +(1,0)  |- (s7.east);
\draw [arrow] (s9) -- (s10);
\draw [arrow] (s10) -- (d2);
\draw [arrow] (d2) -- node[anchor=north west] {yes} (end);
\draw [arrow] (d2.west) node[anchor=south ] {no} -|  +(-.75,0) |- (s2);

\end{tikzpicture}
    \caption{Workflow of the methodology.}
    \label{fig:wf}
    \end{figure}
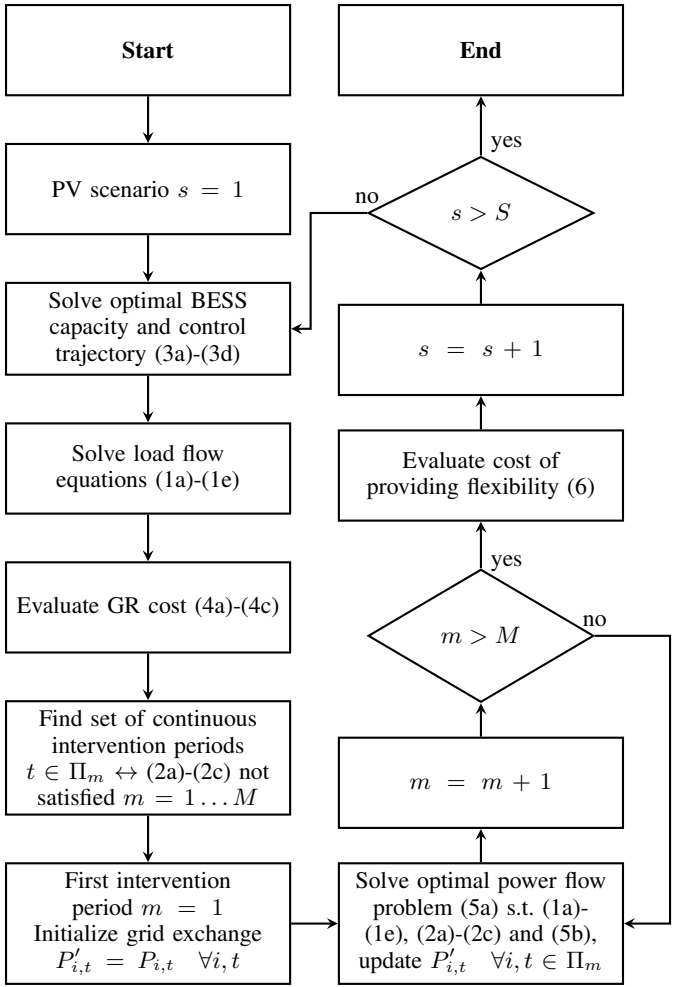

    Finally, the cost of providing flexibility is defined as the difference between the operating costs evaluated after solving \cref{eq:flex_opf_opt} $\opex_i^\prime$ and the original operating cost $\opex_i$ from optimal control trajectory \cref{eq:min_totex}. The methodology is graphically summarised in \cref{fig:wf}. We further distinguish the case where storage is available and can be controlled by the DSO from the case where no storage is available. In the latter case  $\pbat$   is removed from the power balance, and PV curtailment and reactive power injection are the only  options available to the DSO.
	
	\begin{subequations}
	\begin{align}
	\label{eq:min_totex}
	\min_{\ebatcap,\pbat}& \opex + \sigma +  R \cdot \left(\cfbat + \cbat \cdot \ebatcap \right)\\
	\label{eq:opex0}
      \opex &=  \sum_{t=1}^T \left( P^{-}_{t} \cdot \cimp_t - P^{+}_{t} \cdot \cexp_t \right) \\
	\label{eq:sigma}
	\sigma &= \sum_{t=1}^T P^{+\tf{bat}}\cdot 10^{-6}\\
	\label{eq:bat_ann}
    R &= \frac{r \cdot (1 + r) ^{L}}{ (1 + r) ^ {L} - 1}
	\end{align}
	\end{subequations}
	
	\begin{subequations}
    \begin{align}
    \label{eq:reinf}
      C_\text{reinf} &= R^\tf{grid} \cdot \left( C_\text{reinf,line} + C_\text{reinf,trafo} \right)\\
      \label{eq:reinf_line}
     C_\text{reinf,line} &= \sum_{(l) \in \mathbb{L}} \delta_l \cdot c^\tf{line} \cdot d_l\\
     \label{eq:reinf_trafo}
     C_\text{reinf,trafo} &=  \delta_{tr}  \cdot c^\tf{trafo} \cdot \left(\max_t  S^\tf{tr}_{t}\right) \\
     \label{eq:bin_line}
     \delta_l &=  \begin{cases}
                    1 & \text{if } \max_t  I_{l,t} >I^\tf{max} \\
                    0 & \text{otherwise}
                \end{cases}\\
    \label{eq:bin_trafo}
     \delta_{tr} &= \begin{cases}
                    1 & \text{if } \max_t  S^\tf{tr}_{t} >S^\tf{tr,max} \\
                    0 & \text{otherwise}
                \end{cases}
    \end{align}
    \end{subequations}

    \begin{subequations}
    \begin{gather}
        \label{eq:flex_opf_opt}
    \min_{\pcur_{i,t},\pbat,Q^\tf{\pv}_{i,t}}    \qquad \sum_{i\in \mathbb{S},t \in \Pi_m } \pcur_{i,t}\\
    \label{eq:soc_constr}
	E^{\prime,\tf{bat}}_{i,t} =E^\tf{bat}_{i,t}  \quad \text{for } t=\left\{ \underline{\Pi}_m,\overline{\Pi}_m\right\} \quad i\in\mathbb{S}\\
	\label{eq:q_max}
	-q_r^\tf{max}\cdot P^\pv_{\tf{cap},i} \leq Q^\pv_\sit \leq q_r^\tf{max}\cdot P^\pv_{\tf{cap},i}
    \end{gather}
    \end{subequations}

    \begin{equation}
    \label{eq:flex_cost}
    \Delta \opex = \sum_i \opex^\prime_i- \opex_i
    \end{equation}

	 \section{Case study}
	
\begin{figure*}
    \centering
    \includegraphics[width=.7\textwidth]{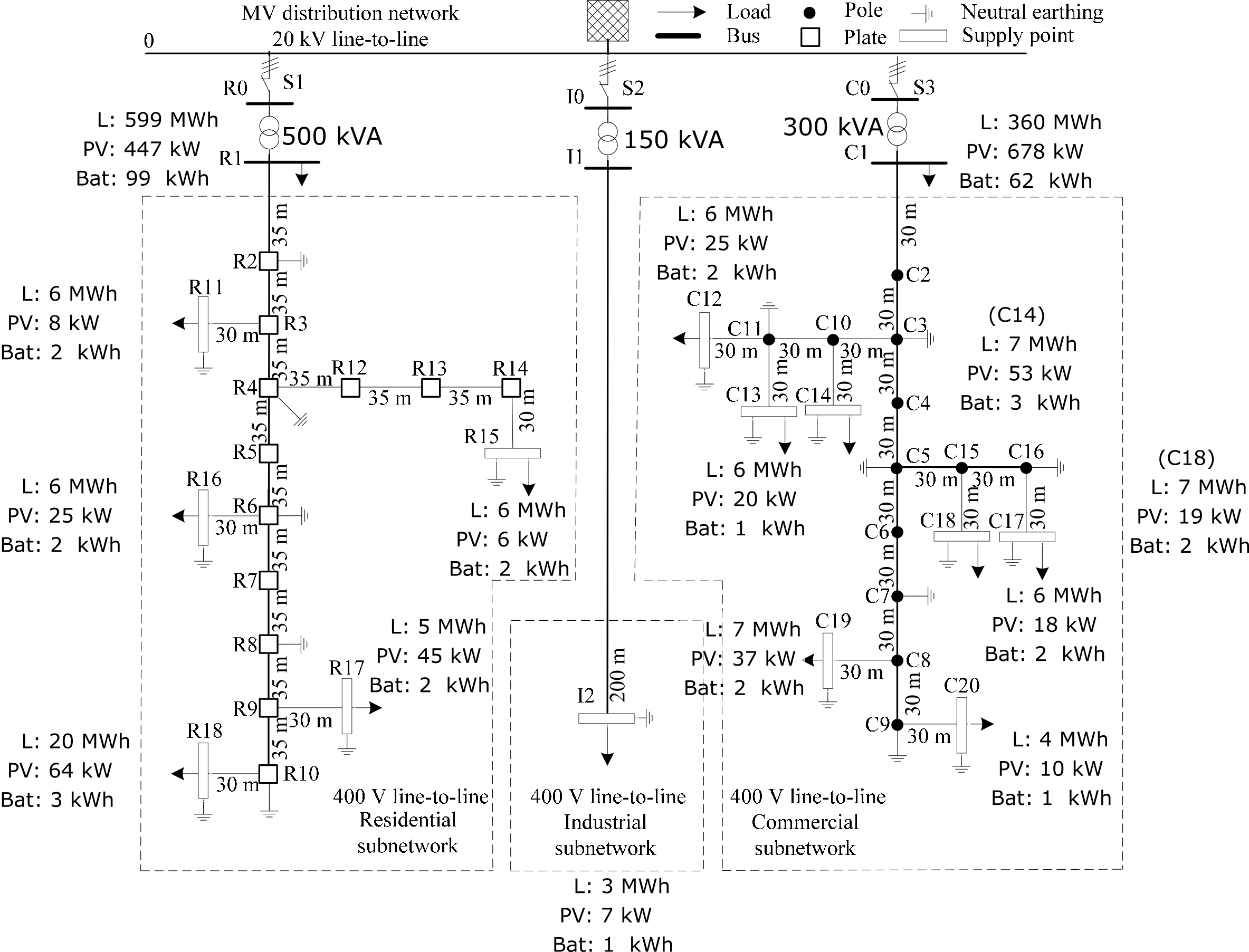}
    \caption{Illustration of the \textit{CIGRE} low-voltage network adapted from \cite{CIGRE2009}.}
    \label{fig:cigre_net}
    \end{figure*}

	 The \textit{CIGRE} low-voltage network \cite{CIGRE2009}, depicted in \cref{fig:cigre_net}, serves as our test case. Loads and PV generation profiles cover one year at 15-\,min resolution. The load profiles come from an internal load database acquired over several projects \cite{flexi,flexi2}. They have been allocated to each network bus, minimising the sum of the differences between the \textit{CIGRE} loads' apparent power and the \nth{99} quantile of the internal load profiles. Similarly, the roofs' characteristics that give the PV potential capacity are drawn from a building database from the Swiss building registry\footnote{\url{https://www.housing-stat.ch/fr/accueil.html}}, minimising the sum of the differences between the annual allocated electricity demand and the building estimated electricity demand. The  annual energy demand and maximum PV potential capacity for each prosumer are listed in \cref{tab:flex_system_cap}. The total annual energy demand and PV capacities are 1050\,MWh and 1500\,kW, respectively. 
	 
	 Each prosumer's BESS capacity and control trajectory is optimised using a time-of-use tariff reported in  \cref{tab:flex_tarr}. The import rate is 23.92 \ctskwh during peak hours and 15.16 \ctskwh during off-peak hours. The export rate is 8.16 \ctskwh all the time. The cost of a BESS $\cbat$ is assumed to be 182 \unit{\frac{CHF}{kWh}}, with a fixed component $\cfbat=0$, as reported in \cref{tab:flex_params}. The BESS capacities reported in \cref{tab:flex_system_cap} correspond to the optimised capacities at the maximum PV capacity.
	  
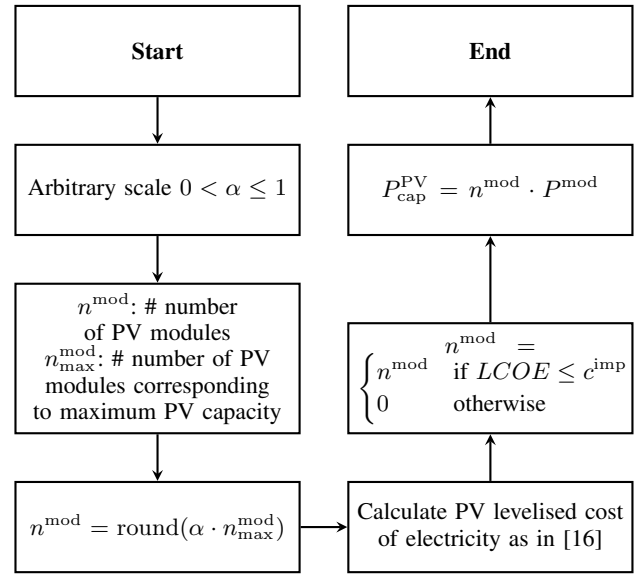
\begin{figure}[h]
    \centering
    \newcommand{\bwidth}{.4\columnwidth}
\newcommand{\vnodespace}{.05\columnwidth}
\newcommand{\vvnodespace}{.07\columnwidth}
\newcommand{\hnodespace}{.07\columnwidth}
\newcommand{\bheight}{1.2cm}

\tikzstyle{box} = [rectangle, draw=black, thick,minimum width=\bwidth, minimum height=\bheight,text centered,text width=\bwidth]
\tikzstyle{decision} = [diamond, draw,minimum width=3cm, minimum height=1cm, text centered, draw=black, thick,fill=white]
\tikzstyle{arrow} = [thick,->,>=stealth]

\begin{tikzpicture}[node distance=.15\columnwidth,auto,font=\small]

\node (start) [box]  {\textbf{Start}};
\node (s0) [box,below=\vvnodespace of start] {Arbitrary scale $0<\alpha \leq 1$};
\node (end) [box,right=\hnodespace of start] {
\textbf{End}};

\node (s1) [box,below=\vvnodespace of s0] {$n^\tf{mod}$: \# number of PV modules\\
$n^\tf{mod}_{\max}$: \# number of PV modules corresponding to maximum PV capacity
};

\node (s2) [box,below=\vvnodespace of s1] {$n^\tf{mod} = \tf{round}(\alpha\cdot n^\tf{mod}_{\max})$};

\node (s3) [box,right=\hnodespace of s2] {Calculate PV levelised cost of electricity as in \cite{Lai2017}};

\node (s4) [box,above=\vvnodespace of s3] {$n^\tf{mod}=\begin{cases}
                     n^\tf{mod} & \text{if } LCOE\leq c^\tf{imp}\\
                     0 &  \text{otherwise} \end{cases}$};
                
\node (s5) [box, below=\vvnodespace of end] {$P_\tf{cap}^\tf{PV}=n^\tf{mod}\cdot P^\tf{mod}$};

\draw [arrow] (start) -- (s0);
\draw [arrow] (s0) -- (s1);
\draw [arrow] (s1) -- (s2);
\draw [arrow] (s2) -- (s3);
\draw [arrow] (s3) -- (s4);
\draw [arrow] (s4) -- (s5);
\draw [arrow] (s5) -- (end);
\end{tikzpicture}
    \caption{Determination of the installed PV capacity for a particular system.}
    \label{fig:pv_scale_wf}
    \end{figure}

\begin{figure}[!hb]
    \centerline{\includegraphics[width=\fw]{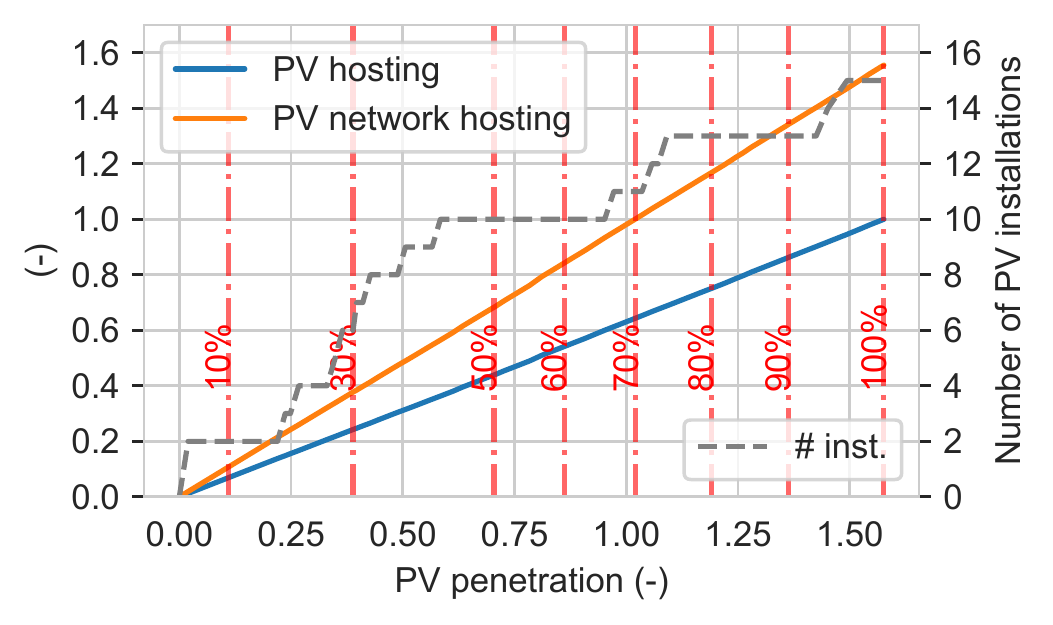}}
    \caption{PV hosting and PV network hosting ratio (left axis) and the number of installed systems (right axis, maximum is 15 systems). The red dashed lines are the selected \pv penetration scenarios. The numbers indicated in red are the scale used to vary the \pv capacity.}
    \label{fig:pv_scale}
    \end{figure}

	 The number of modules giving the maximum potential PV capacity for each system is scaled by an arbitrary number (reported in red in \cref{fig:pv_scale}) to construct various PV penetration scenarios. The number of modules is then rounded to the nearest integer. To prevent an unrealistically small installation, we assume that if the resulting PV levelised cost of electricity is greater than a threshold of 23.92 \ctskwh (corresponding to the high import rate), the resulting PV capacity is set to zero for this scenario. The PV levelised cost of electricity is calculated assuming a PV lifetime of 25 years, a discount rate of 3\%, a PV cost of 83\,\unit{\frac{cts}{W}}, and a fixed cost of 10'050 CHF. This process is graphically summarised in \cref{fig:pv_scale_wf}. The PV and battery capacity indicated in \cref{fig:cigre_net,tab:flex_system_cap} corresponds to the maximum PV penetration scenario (scale = 100\%). The global horizontal and diffuse irradiance data are extracted from a weather station in Pully \footnote{Data available at \url{https://gate.meteoswiss.ch/idaweb}}. 
	 
	 We assumed the costs for GR of the transformer and lines are 60\kchfmva and 70\,\unit{\frac{kCHF}{km}}.  The transformer's and lines' lifetime is assumed to be 30 years, as reported in \cref{tab:flex_params}. 
	 
        \begin{table}[!ht]
        \centering
        \caption{Systems data}
        \label{tab:flex_system_cap}
       \begin{tabular}{lccc}
        \toprule
              &  \textbf{Annual demand} &  \textbf{PV max capacity}&  \textbf{Battery capacity} \\
              & (MWh)           & (kW)                  & (kWh)\\
        \midrule
    R1 &        598.97 &                447.09 &                  141.13 \\
    R11 &          6.09 &                  8.19 &                    2.89 \\
    R15 &          6.13 &                  6.30 &                    2.18 \\
    R16 &          5.58 &                 24.89 &                    2.48 \\
    R17 &          5.29 &                 44.74 &                    2.30 \\
    R18 &         19.82 &                 64.27 &                    4.71 \\
    I2 &          2.88 &                  6.93 &                    1.21 \\
    C1 &        360.34 &                678.35 &                   87.96 \\
    C12 &          5.89 &                 19.53 &                    1.39 \\
    C13 &         11.73 &                 40.64 &                    4.83 \\
    C14 &          7.02 &                 53.25 &                    3.69 \\
    C17 &          5.84 &                 17.64 &                    2.37 \\
    C18 &          6.56 &                 18.90 &                    3.10 \\
    C19 &          7.34 &                 36.86 &                    2.65 \\
    C20 &          3.59 &                 10.08 &                    1.09 \\
    \midrule
    Total &       1053.08 &               1477.69 &                  263.97 \\
        \bottomrule
        \end{tabular}
    \end{table}
    
    \begin{table}[!ht]
    \centering
    \caption{Tariff}
    \label{tab:flex_tarr}
    \begin{tabular}{lll}
    \toprule
            &  \textbf{Hours} & \textbf{Tariff (cts/kWh)} \\
    \midrule
    \multirow{3}{*}{\cimp} & Mon-Fri 06h-22h   & 23.92\\
                            & Mon-Fri 22h-06h   & 15.16\\
                            & Sat-Sun all-day   & 15.16\\
    \midrule
    \cexp                  &                   & 8.16\\
    \bottomrule
    \end{tabular}
    \end{table}
    
    \begin{table}[!ht]
        \centering
        \caption{System, network, and battery parameters}
        \label{tab:flex_params}
        \begin{tabular}{@{}llcc}
        \toprule
         & \textbf{Param.}      & \textbf{ Unit} & \textbf{Value}\\
         \midrule
         & $T$                  & -         & 35040 \\
         & $\ts$                & s         & 900   \\
         & $L$                  & years     & 9     \\
         & $r$                  & -         & 0.03  \\
         \midrule
         &  $\underline{V},\overline{V}$& pu& 0.95-1.05 \\
         & $\overline{I}$ & kA & 1\\
         & $\overline{S}^\tf{tr}$ & kVA & see \cref{fig:cigre_net}\\
         & $c^\tf{trafo}$       & kCHF/MVA   & 60    \\
         & $c^\tf{line}$        & kCHF/km   & 70    \\
         & $r^\tf{grid}$        & -         & 0.03  \\
         & $L^\tf{grid}$        & years     & 30    \\ 
         \midrule
         & $\cbat$              & CHF/kWh   & 182   \\
         & $\cfbat$             & CHF       & 0     \\
         & $q_r^\tf{max}$      & MVAr/MW  & 0.4     \\
         \bottomrule
        \end{tabular}
        
    \end{table}

	\section{Results}
	
\begin{figure}[!ht]
		\newcommand{\fhe}{3.5cm}
        \centering
        \subfloat[\label{fig:trafo_pr}Transformer loading level and bus voltage.]{\includegraphics[width=\fw]{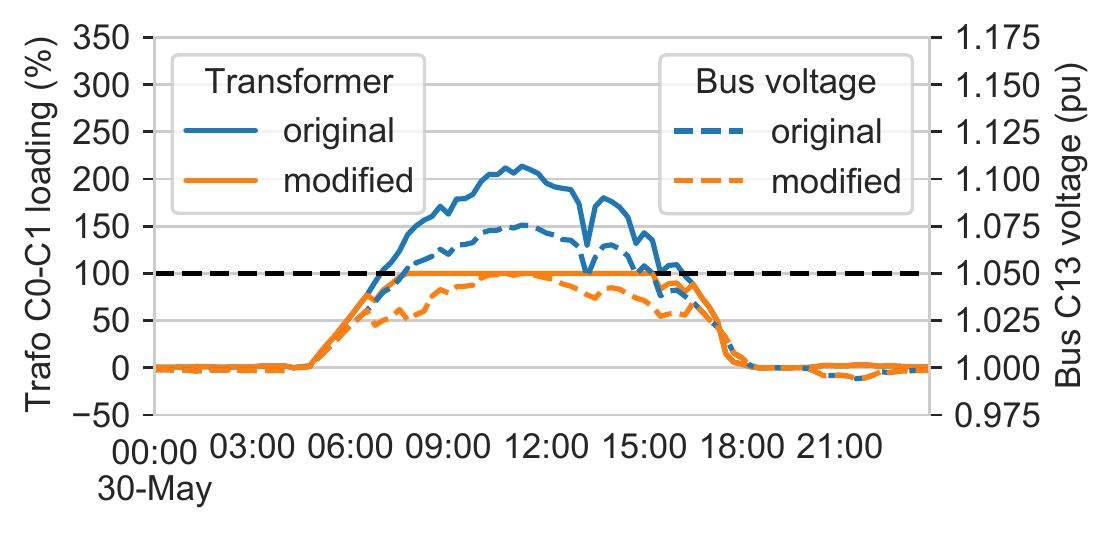}}\\
        \vspace{0.5cm}
        \includegraphics[width=\fw]{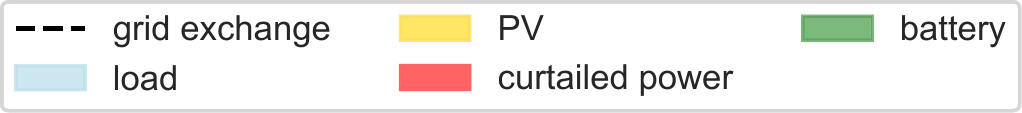}\\
        \vspace{-0.3cm}
        \subfloat[\label{fig:syst_profile_0} Original system operation.]{\includegraphics[height=\fhe]{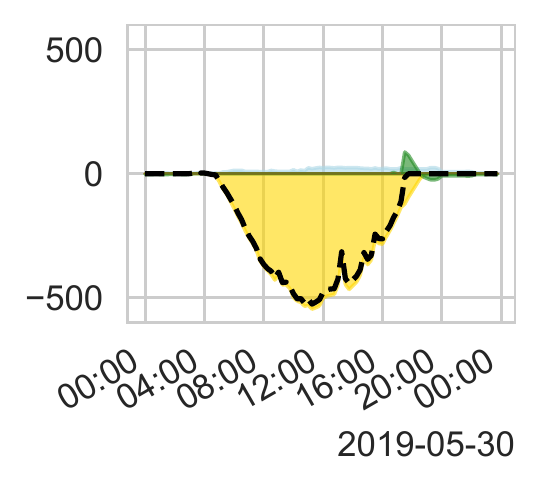}} 
        \subfloat[\label{fig:syst_profile_1} Modified system operation.]{\includegraphics[height=\fhe]{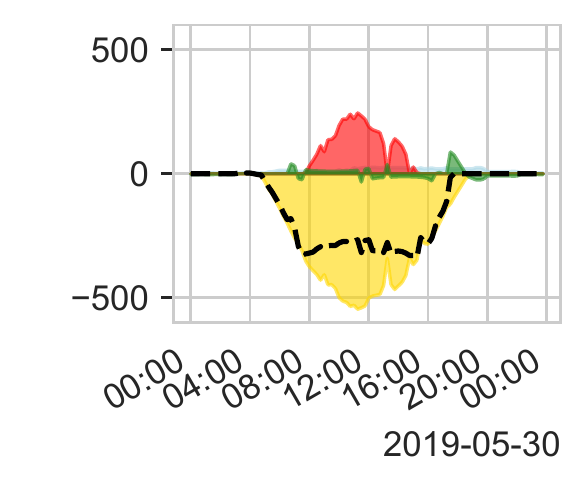}} \\
        \subfloat[\label{fig:soc_profile}State of  charge evolution during a day.]{ \includegraphics[width=\fw]{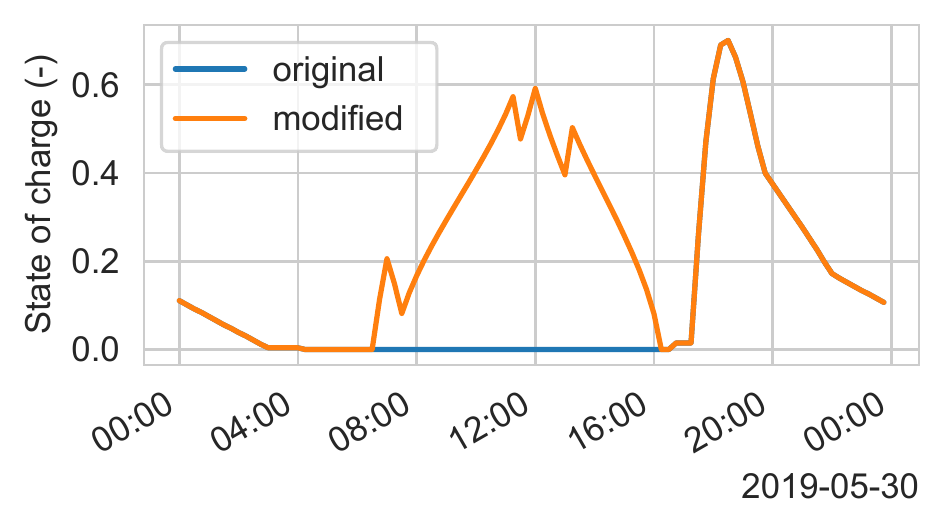}}
        \caption{Illustration of the network and one system active power for one day, for the 100\% PV capacity scenario.}
        \label{fig:profile}
    \end{figure}

	\Cref{fig:profile} illustrates a typical operating day of the maximum PV penetration scenario. As the total PV generation increases, the C0-C1 transformer's apparent power exceeds the rated capacity (transformer loading above 100\% in \cref{fig:trafo_pr}). Similarly, the voltage level exceeds 1.05 pu in bus C13. This is caused by an excess of PV energy, as illustrated in \cref{fig:syst_profile_0} for the system located in bus C1. To resolve such problems, a significant fraction of the PV energy needs to be curtailed, and the battery power profile is modified compared to its original trajectory (see \cref{fig:syst_profile_1} for the same system). The corresponding battery state of charge trajectory is kept untouched except during the intervention period, as depicted in \cref{fig:soc_profile}.  
	
\begin{figure}[!ht]
        \centering
        \includegraphics[width=\fw]{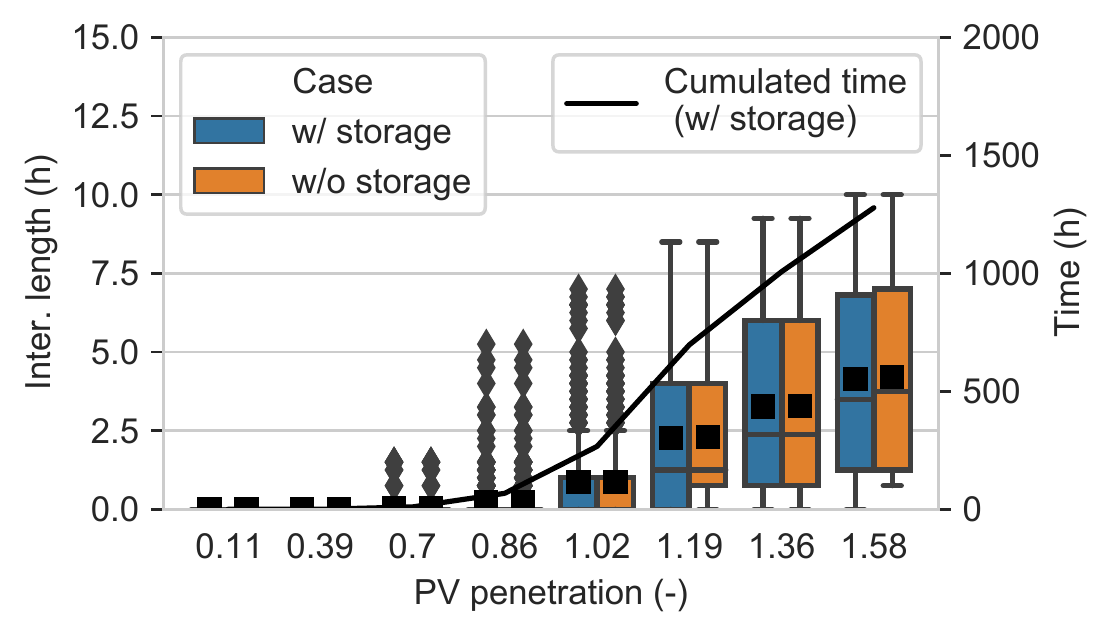}
        \caption{Duration distribution (left axis). Averages are indicated with a square and outliers with a diamond. The total intervention time is on the right axis.}
        \label{fig:interv_time}
    \end{figure}

	In the previous example, the intervention period is about 10 hours long. \Cref{fig:interv_time} shows the distribution of the duration of the intervention periods as well as the total duration of the interventions. As PV penetration increases, the intervention duration tends to increase. The total duration of the interventions increases to about 1300 hours (4 hours per day on average). No significant difference is observed between the case with storage and the case without storage. \cref{tab:net_viol_summary} reports the number of hours when the transformer experiences overloading and buses experience overvoltage. No line overloading is observed. Transformer overloading is observed starting at a PV penetration of 70\%. It occurs for 5 hours only. Bus overvoltage occurs starting at  a PV penetration of 119\% and the cumulative time is about 30 hours. The differences between the case with storage and the case without storage are insignificant. These results highlight that, for this network, the limiting component  is the transformer loading. This can be easily solved by curtailing PV energy.
	
	As reported in \cref{tab:curt}, an insignificant fraction of the total PV energy needs to be curtailed at a PV penetration of 70\%. At the maximum PV penetration and considering no storage, 8.4\% of the 1660 MWh needs to be curtailed. Using storage reduces this quantity to 2.9\%. The amount of curtailed PV energy impacts the financial performance of the systems. 
	
\begin{figure}[!ht]
        \centering
        \includegraphics[width=\fw]{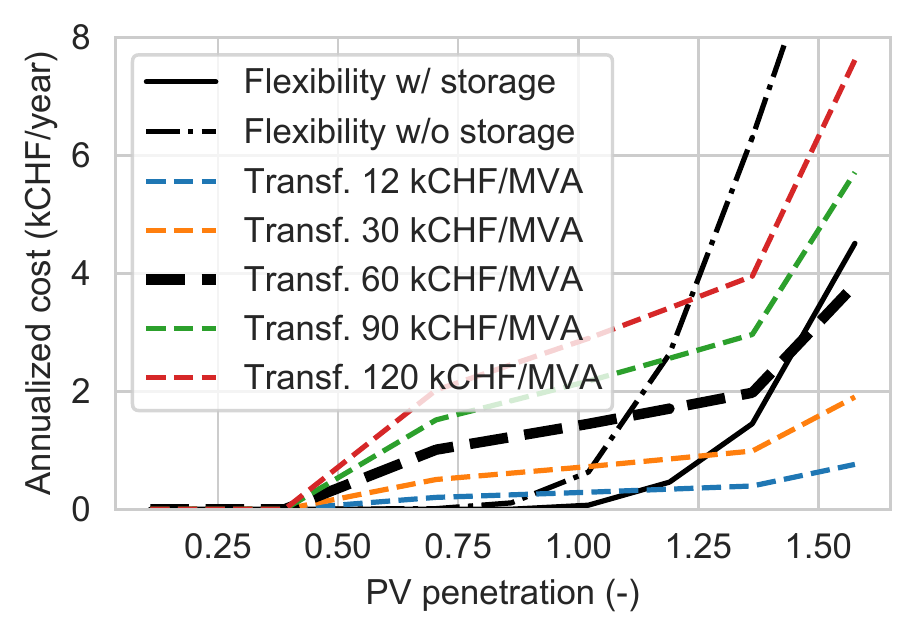}
        \caption{Annual cost of providing flexibility versus grid reinforcement.}
        \label{fig:flex_cost}
    \end{figure}

	The cost of providing flexibility is reported in \cref{fig:flex_cost}. As it reduces the amount of curtailed PV energy, storage significantly reduces the cost of flexibility. This cost has to be compared with the GR cost. As no line-overloading is observed, we perform a sensitivity analysis on the transformer specific cost around its nominal value (60\kchfmva). For an optimistic case where the transformers cost 12 \kchfmva, using distributed flexibility is profitable up to a PV penetration of 110\%. This break-even point moves to 145\% for a 60\kchfmva transformer. Above such a price, it is always profitable to utilise distributed flexibility for all PV penetrations.

	\begin{table}
        \centering
        \caption{Summary of the network violations for the case with storage. Any differences between the case without and with storage are indicated between brackets.}
        \label{tab:net_viol_summary}
        
        \begin{tabular}{llrr|rr}
        \toprule
         &       & \multicolumn{2}{r|}{\textbf{Transformer overloading}} & \multicolumn{2}{l}{\textbf{Bus overvoltage}}             \\
         &       & \textbf{time (h)}   &\textbf{ max  (\%)}                      & \textbf{time (h)}         &\textbf{ max  (pu)} \\ 
         \midrule
         \parbox[t]{1mm}{\multirow{8}{*}{\rotatebox[origin=c]{90}{\textbf{\pv penetration}}}}
         & 11\%  & -          & -                              & -                & -                             \\
         & 39\%  & -          & -                              & -                & -                             \\
         & 70\%  & 5       & 110                            & -                & -                             \\
         & 86\%  & 44      & 134                            & -                & -                             \\
         & 102\% & 173(+2)    & 158                            & -                & -                             \\
         & 119\% & 523(+7)    & 186                            & 29(+1)           & 1.06                          \\
         & 136\% & 799(+5)    & 215                            & 439(+3)          & 1.07                          \\
         & 158\% & 1044(+16)  & 246                            & 846(+6)          & 1.09                          \\ 
         \bottomrule
        \end{tabular}

        
    \end{table}

    \begin{table}
        \centering
        \caption{Energy curtailed and \pv production}
        \label{tab:curt}
        \begin{tabular}{llccc}
        \toprule
        &  & \multicolumn{2}{c}{\textbf{Energy curtailed (\%)}} &  \textbf{Generation (MWh)} \\
              &  &  \textbf{w/ storage} &  \textbf{w/o storage} & \\
        \midrule
        \parbox[t]{1mm}{\multirow{8}{*}{\rotatebox[origin=c]{90}{\textbf{\pv penetration}}}} &
                    11\% &         0.0 &          0 &          115 \\
                &    39\% &         0.0 &          0.0 &          409 \\
                 &   70\% &         0.0 &          0.0 &          742 \\
                  &  86\% &         0.0 &          0.1 &          907 \\
                 &   102\% &         0.0 &          0.8 &         1076 \\
                &    119\% &         0.3 &          2.6 &         1253 \\
                &    136\% &         0.9 &          5.4 &         1435 \\
                &    158\% &         2.9 &          8.4 &         1660 \\
        \bottomrule
        \end{tabular}
    \end{table}

	\section{Discussion}
	The results highlight that distributed flexibility is profitable over GR for a significant range of PV penetrations. Several underlying assumptions need to be discussed. First, both PV and storage investment costs are assumed to be undertaken by the prosumers. We do not tackle the question of the optimal sizing of BESS and PV systems as in \cite{pscc} but fix the PV capacity to generate a realistic PV penetration scenario and only solve the optimal BESS capacity problem instead. At this stage, grid support activities like PV curtailment and deviation of the BESS's optimal control trajectory are not considered as future revenue. Thus, we assume that the DSO itself does not undertake the risk of investing in distributed storage. This discards the question of the profitability of BESSs for such a grid support application as addressed by \cite{Gupta2021,Gupta2021a}. 
	
	A second strong assumption is buried in the BESS's optimal control trajectory. Such optimisation problems assume a perfect forecast of the uncontrollable load and PV generation. Similarly, the optimal power-flow problem also relies on a similar assumption. To turn such a concept into practice, there is a need for dedicated control algorithms that take load and generation uncertainties into account and that minimise future impacts on individual system operation. 
	
	Third, the costs of providing flexibility and GR are compared by annualising the operating and investment costs. This implies that the system  would operate identically for the following 30 years without considering the future evolution of the technology cost and PV deployment. Our approach must be seen as an evaluation tool to estimate where GR should be undertaken, typically in networks with a high PV penetration potential. Future work may repeat such a study for a wider range of networks with different topologies and load characteristics and extract insights to better understand the profitability of distributed flexibility. 
	
	Finally, we neglected the additional cost of remote control of the prosumers. Those costs can be covered by the savings the DSO makes from using distributed flexibility instead of GR. From another perspective, those savings could be redistributed to the prosumers as a reward for contributing to the network flexibility. As a potential approach, one could define the flexibility capacity $P^\tf{flex}_\tf{cap}$ as the sum of the PV and BESS's power capacity (curtailment and charging can be seen as virtual loads). The difference between the GR and  the cost of flexibility in \cref{fig:flex_cost} corresponds to the DSO's savings, which could be redistributed to prosumers according to their flexibility capacity. The value of the flexibility capacity could be defined as: 
	
	\begin{equation}
	    C^\tf{flex,value} = \frac{C_\tf{reinf}-\Delta \opex}{R^\tf{grid}}\cdot \frac{1}{\sum_i P^\tf{flex}_\tf{cap}}
	\end{equation}
	
\begin{figure}[!ht]
	    \centering
	    \includegraphics[width=\fw]{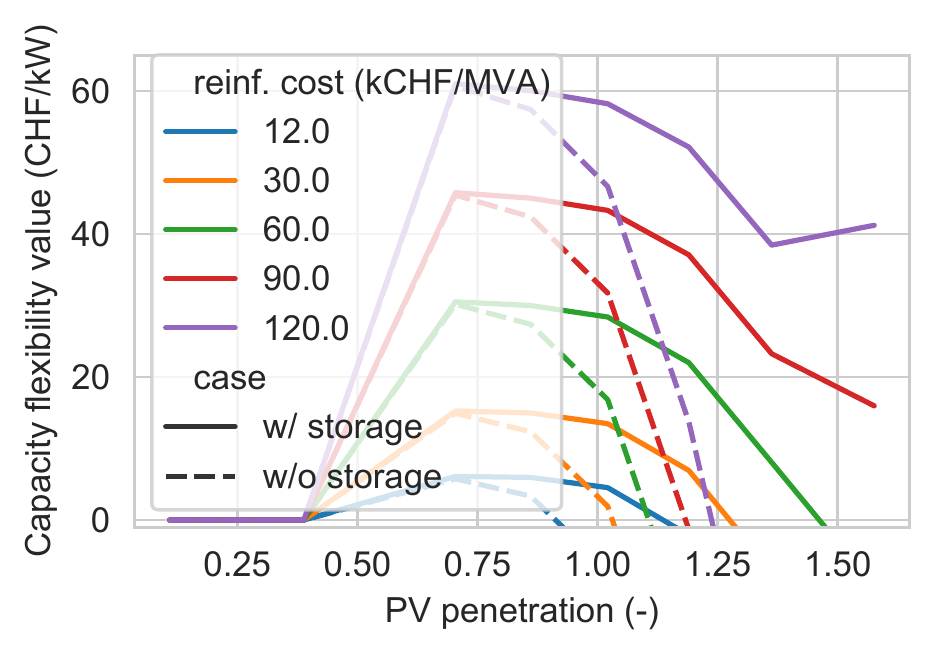}
	    \caption{Flexibility capacity value.}
	    \label{fig:flex_cap_val}
	\end{figure}

	This quantity can be seen as a one-time subsidy to the prosumer's system. The value of the flexibility capacity is reported in \cref{fig:flex_cap_val}. The flexibility capacity value decreases more significantly without storage. For a typical reinforcement cost of 60 \kchfmva, such value lies in 20-30 CHF/kW, which corresponds to about 5-10\% of the 420 CHF/kW Swiss PV one-time subsidy\footnote{Appendix 2.1, ch2.3 of OEnER \url{https://www.fedlex.admin.ch/eli/cc/2017/766/fr\#lvl_d1384e202/lvl_d1384e203/lvl_2}}. 
	
	\section{Conclusion}
	This work aims to complement the recent literature on how to price and reward flexibility. We presented an approach to assess the cost of distributed flexibility and grid reinforcement in a low-voltage network for several PV penetration levels. Our approach starts with optimising the BESS's capacity and optimal control trajectory for each prosumer located in the network. The PV capacity is fixed to control the PV penetration at the network scale but could be included in the optimisation problem. The second step is to solve the load flow equations to evaluate potential violations of network constraints  (bus overvoltage, line ampacity violation, and transformer overloading). At this stage, we evaluate the cost of grid reinforcement. Then for each continuous period when violations of constraints  were observed, we solved an optimisation problem aiming to minimise the amount of curtailed PV energy. This mimics an action undertaken by the DSO to keep the network safe. As this action is a deviation of the prosumers' optimal control trajectory, the associated cost overrun is paid to the prosumers. These cost overruns are considered to be the cost of providing flexibility. We compared the cost of grid reinforcement and providing flexibility. Distributed flexibility is more profitable than grid reinforcement when the need for grid reinforcement arises. We showed that a BESS significantly lowers the cost of providing flexibility compared to considering only PV curtailment.  The BESS's profitability is ensured by the definition of the objective function of the first-stage optimisation, which minimises the prosumer's energy bill.  Such a scheme transfers the risk of investing in a BESS from the DSO to the prosumers, assuming a sufficient incentive, e.g. an electricity tariff, exists to encourage the latter to invest in such technology. Future work may consider more-advanced electricity tariffs and evaluate how a DSO could select the most appropriate tariff to encourage investment in flexibility capacity and to lower the distributed flexibility cost. We also discussed a fair approach to reward flexibility as a one-time contribution to the prosumer's system. Finally, we showed that distributed flexibility is more profitable than grid reinforcement for PV penetration up to 110-145\%.
	
	Future work should focus on control algorithms taking into account load and PV generation uncertainties. Possible deployment schemes for a practical application should also be investigated.

	\newpage
	
	
	\bibliographystyle{IEEEtran}
	\bibliography{IEEEabrv,biblio}
	%
		


\end{document}